# Overfitting the literature to one set of stimuli and data

Tijl Grootswagers[1,2,3], Amanda K. Robinson[3]

[1]The MARCS Institute for Brain, Behaviour and Development, Sydney, Australia

[2]Western Sydney University, Sydney, Australia

[3]School of Psychology, University of Sydney, Sydney, Australia

## Abstract

The fast-growing field of Computational Cognitive Neuroscience is on track to meet its first crisis. A large number of papers in this nascent field are developing and testing novel analysis methods using the same stimuli and neuroimaging datasets. Publication bias and confirmatory exploration will result in overfitting to the limited available data. The field urgently needs to collect more good quality open neuroimaging data using a variety of experimental stimuli, to test the generalisability of current published results, and allow for more robust results in future work.

## Background

How many ways are there to look at one set of data? In a highly influential paper, (Kriegeskorte et al., 2008b) compared human fMRI responses to electrophysiological recordings from Monkey inferior temporal cortex (obtained from Kiani et al., 2007), revealing a striking similarity in the representations of objects between species (Kriegeskorte et al., 2008b). In addition, this study introduced Representation Similarity Analysis (RSA) framework (Kriegeskorte et al., 2008a) to compare information representations between fMRI and electrophysiological recordings. This framework is now a widely used method for comparing information across modalities (Kriegeskorte and Kievit, 2013). The RSA framework was later used in a landmark study that used MEG and fMRI to track object representations in space and time (Cichy et al., 2014). For ease of comparing their results, this study used the same stimulus set (Figure 1) as Kriegeskorte et al. (2008). So far so good, but these stimuli and corresponding fMRI and MEG data have now formed the basis for over 30 (and counting) publications (Figure 2). These studies have yielded important information about how new analysis methods can be used to give insight into the visual system. Yet, it is undeniable that overuse of the same stimuli and data will eventually lead to a bias in the literature. A major factor to consider when designing a study is how generalisable the findings will be. Any one study is characterised by details (and limitations) of the experimental design, data collection and analyses. Analysing the same sets of data in different ways or using the same stimulus sets will lead to over-representation and over-generalisation of experiment-specific trends. The intention of this commentary is not to undermine or refute any of these studies, but rather to point out that the field needs to diversify.

## A problematic stimulus set

The stimulus set described above consists of 92 segmented visual objects of animals, people, places and things (Figure 1). At first glance, there is nothing wrong with the set itself. To study object representations, we need stimuli, but controlling for all possible covariates in a set of stimuli is challenging, thus no set is perfect. In fact, the 92 objects were a considerable advance over previous work that had stronger limits imposed by the experimental designs. However, there are still issues with this set, as highlighted in Figure 1. Some stimuli were reported to be ambiguous (Kriegeskorte et al., 2008b). Some do not, on close inspection, belong in the manually specified categories, for example an image of hair is classified as "human body". There are also clear categorical differences between simple image features that covary with the imposed category structure. For example, many animate stimuli contain faces, which on average are visually more similar than stimuli within the inanimate category (Figure 1). These reliable visual similarities likely lead to large-scale pattern differences in neuroimaging data, which could account for the strong animate-inanimate distinction that is often observed in studies that used the 92-object stimuli (e.g., Carlson et al., 2013; Cichy et al., 2014; Kaneshiro et al., 2015; Kriegeskorte et al., 2008b), and is less prominent in studies that used stimulus sets that controlled for systematic visual differences (e.g., Bracci and Op de Beeck, 2016; Grootswagers et al., 2019; Long et al., 2018; Proklova et al., 2019, 2016; Rice et al., 2014).

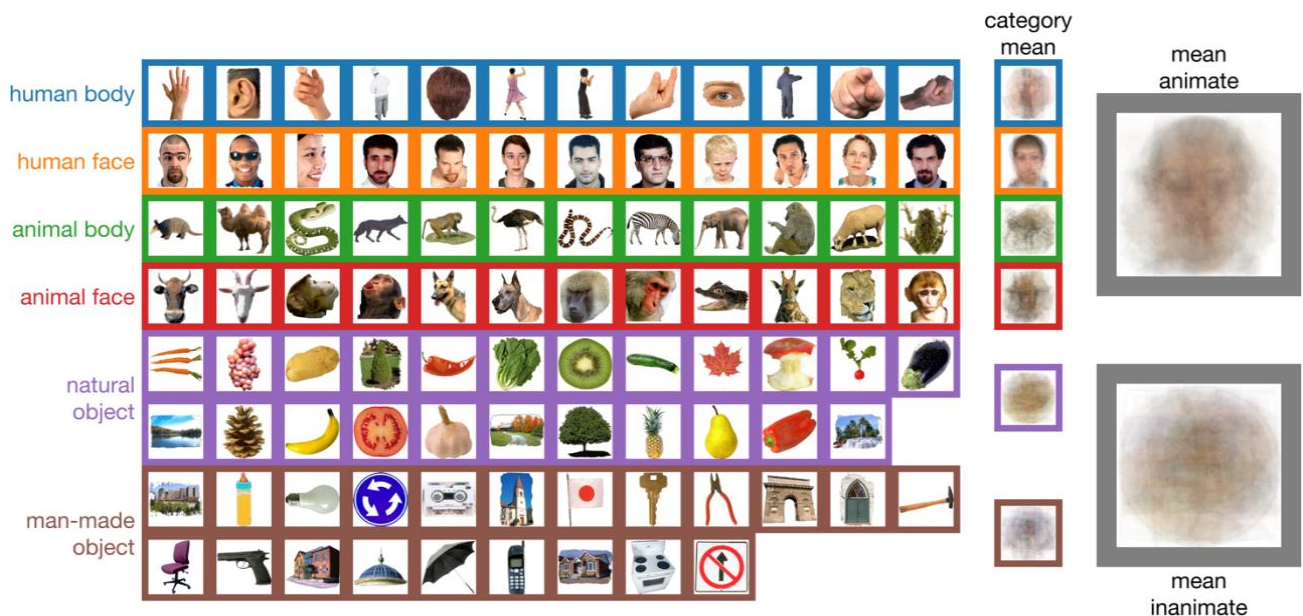

Figure 1. The 92-object stimulus set used in over 35 computational cognitive neuroscience papers. Some of the objects in the 'animate' category are not animate (e.g., hair; row 1 column 5, or a cut-out wolf figurine; row 3 column 4). Some inanimate images are not objects, but rather scenes (e.g., row 6/7 column 1). This stimulus set is often used to highlight a strong animate/inanimate dichotomy in human brain responses, but the categories have consistent visual differences (the rightmost two columns show the means of all images in a category).

These limitations would not constitute a huge problem on their own. They could be addressed in follow-up work that replicates results using a different stimulus set, or a set that specifically controls for the issues above. Indeed, several studies have used variations or entirely different sets to highlight contrasting and complementary findings (For a recent review, see Wardle and Baker, 2020). However, the problem arises because a large amount of published work has used the exact same stimulus set. This leads to a wide-spread issue of generalisability. In addition, the current academic landscape encourages only publishing positive results (Ioannidis et al., 2014), which could mean that we are getting a skewed picture of published results that are specific to the 92-object stimulus set.

OVERFITTING TO ONE NEUROIMAGING DATASET

Overusing the same stimuli is certainly an issue, but it is arguably more worrying that so many studies also use the exact same data (Figure 2). The MEG and fMRI responses to the 92-object stimulus set were made publicly available (Cichy et al., 2016, 2014). This is a gold standard open science practice, and the dataset has certainly been useful for the field: since its release, a large proportion of experimental papers have used this exact dataset to develop new analyses or models, decide on optimal analysis pipelines, or assess the similarity of the data to other modalities. An unavoidable result, however, is that these new and interesting developments are possibly specific to one particular dataset (note, some have shown that their conclusions were supported by multiple datasets). Looking at the same dataset from many different angles over and over again will lead to several findings that are dataset specific. In other words, there is a risk of overfitting to one set of data. Eventually, this will leave us with a body of literature that does not replicate or generalise to new data, which is a waste of time, money, and other resources.

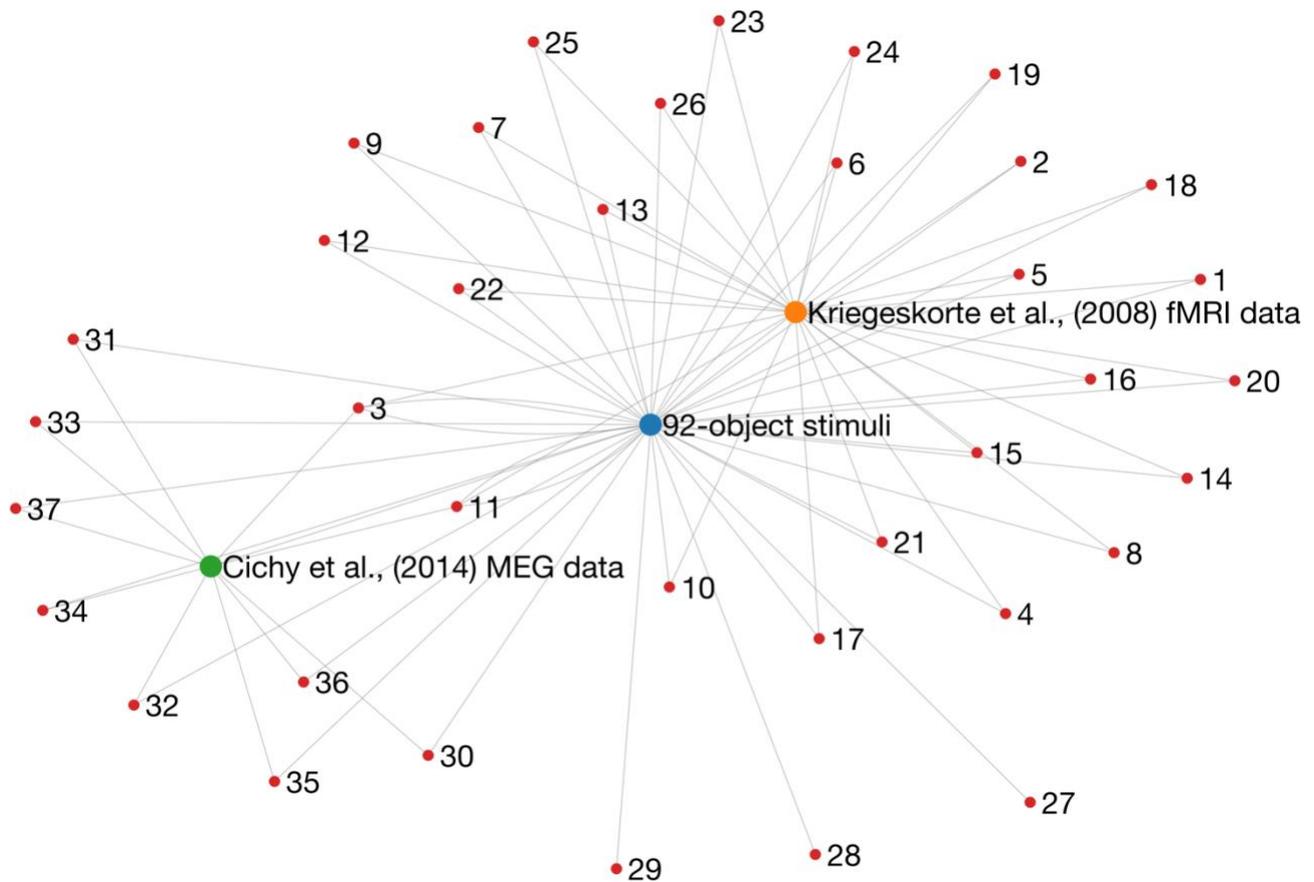

Figure 2. A graph representation of research outputs (nodes) that re-used (edges) the same dataset or stimuli from two influential papers (larger nodes). Importantly, this is not intended to question or refute the findings of any of these studies, but rather to point to a potential issue of generalisability in the literature. Also note that it is likely that this graph is not complete.

We strongly point out that this commentary is not an argument against making data public. On the contrary, if data sharing practices were more common, we may not have had this problem. The current issue lies in the fact that data reuse is very common, but there are very few similar open data sets in this literature. They may exist but are either hard to find, or in a difficult-to-use format. Open data is incredibly useful to validate existing analyses and test new ideas and methods without using immense resources to collect expensive neuroimaging data (resources that do not exist for many researchers). A good balance needs to be struck between collecting new data and reusing existing open data. If resources allow, existing open data could serve as pilot data for new neuroimaging experiments. The resulting output would contribute to greater experimental diversity, yield a new open dataset that can be used in the future and ultimately allow for better generalisability of conclusions.

### THE WAY FORWARD

A large number of papers in Computational Cognitive Neuroscience have used the same dataset and stimulus set, which raises questions about the generalisability of their influential and exciting results. This problem is ongoing, with many of the papers in Figure 2 published in the last 5 years, and several forthcoming works currently on preprint servers. Yet, there are promising signs on the horizon.

First, not all work in Computational Cognitive Neuroscience has relied on these data and a large number of studies have not used the data or stimuli discussed here. Many studies have collected new stimuli and data, and others have re-used different stimuli and datasets. Second, efforts to develop large-scale, systematically selected stimulus databases are a huge step forward, such as THINGS (Hebart et al., 2019), or *ecoset* (Mehrer et al., 2021). These large sets will hopefully will soon be accompanied by high-quality open neuroimaging datasets. Third, data sharing has also become easier through several (free) hosting platforms (e.g., figshare, osf, openneuro), and it is increasingly more common to make data available upon publication. Finally, data formatting standards have been established, such as the brain imaging data structure (BIDS) (Gorgolewski et al., 2016; Holdgraf et al., 2019; Niso et al., 2018; Pernet et al., 2019), which makes it easier to re-use data.

In conclusion, to stave off a replication crisis in Computational Cognitive Neuroscience, we need to strike a delicate balance between taking advantage of existing resources and being aware of the limitations that come with re-using existing data and stimulus sets. While open data will allow the field to keep exploring new ideas without spending huge amounts of public funds or devoting many hours to operating neuroimaging equipment, we equally often need to consider collecting new data to test the reliability of these ideas and improve the body of research as a whole.